# Mobile Microelectrodes: Towards active spatio-temporal control of the electric field and selective cargo assembly

Alicia M Boymelgreen[1], Tov Balli[1], Touvia Miloh[2], Gilad Yossifon[1*]

[1]*Faculty of Mechanical Engineering, Micro- and Nanofluidics Laboratory, Technion – Israel Institute of Technology, Haifa 32000, Israel*

[2]*School of Mechanical Engineering, University of Tel-Aviv, Tel-Aviv 69978, Israel*

*\* Corresponding author: yossifon@technion.ac.il*

*With an eye towards next-generation, smart, micro/nanofluidic devices, capable of responding to external stimuli or changes in environment, we demonstrate a means to achieve dynamic control of the spatio-temporal properties of the electric field in a standardized microfluidic chamber. Typical top-down patterning, currently used to design the field distribution, is replaced by freely-suspended colloids which locally disturb the electric field from the bottom-up. Even under uniform forcing, polarization of the colloid induces the formation of strong, three-dimensional gradients at its surface - essentially repurposing it into a portable floating electrode whose precise location can be manipulated to reconfigure the electric field in real time. Focusing on active Janus colloids as a sample platform, we measure the strength of the induced gradients and highlight the advantages of a colloid-based system by revealing a prototype for an all-in-one cargo carrier, capable of on-demand, selective, label-free assembly and transport of micro/nano sized targets.*

The simultaneous shifts towards flexible, "smart" systems[1,2] and bottom-up manufacturing processes have successfully combined to increase precision, decrease characteristic lengths and increase flexibility[3] across a broad range of applications; from structured, reconfigurable materials[4] to electrical circuit components[5–7]. The current work is motivated by the vision of extending these opportunities to micro/nanofluidic systems executing functions such as separation, analysis and bottom-up material assembly.

Focusing on the popular category of electrokinetically driven devices, in which the non-uniformity of the electric field is often a critical element (particularly in systems controlled by AC signals where broken-symmetry enables the attainment of a net effect, despite the zero time average of the field[8,9]), we observe one unifying common denominator; the geometry of chamber and electrodes and thus by extension, the spatial distribution of the field-gradient must be predetermined and optimized before fabrication begins. Accordingly, the field gradients - whether induced by dielectric structures, non-uniform active or floating electrodes - are essentially a fixed, inherent property of each individual system[10].

In contrast, in the current proposal, where portable colloids are added to the suspending fluid, spatio-temporal field properties are determined at the time of operation, enabling the end user to dynamically adjust the field in a standardized microfluidic chamber to complement the conditions within (Figure 1a). Efficiency may be enhanced by optimizing the geometry, material properties and functionalization of the portable colloids[11] with the potential to create complex structures via bottom-up self-assembly[4,12,13].

To demonstrate the underlying principles, we have used symmetry-broken metallodielectric Janus spheres (where one hemisphere is conducting (gold (Au)) and the second dielectric (polystyrene (Ps))) as the colloidal platform for two reasons: the asymmetry results in stronger induced gradients when compared to homogeneous spheres[14] and their propulsion as active colloids[15], whose speed and direction of motion is dependent on the field frequency offers an inbuilt means for manipulation of the spatial distribution of the floating electrodes. Increased precision can be obtained by integrating technologies for directed motion such as the automated control loop recently demonstrated by Mano et al.[16]. At the same time, it is emphasized that induced gradients are not limited to the geometry of Janus sphere but have also been observed in the gap between a homogeneous gold sphere adjacent to a wall[14] and are expected to form in other asymmetric configurations which may be optimized for specific applications. Non-active colloids, may be advantageous in realizing complex 3D floating electrode structures which can be built from the bottom-up[17] using micro/nanopatterning[18,19] and self-assembly[12,13] techniques. A secondary mechanism, such as magnetophoresis[20,21] may be incorporated to manipulate the particle location.

The three-dimensional distribution of the induced field gradients around the Janus particle are modelled using a combination of numerical simulations and experiment. Since direct measurement of the field gradients is impractical, we employ dielectric nanoparticles as sensors to detect their magnitude and distribution. In the presence of an electric field gradient, these nanosensors experience a dielectrophoretic force where depending on the frequency of the applied field, they will be either attracted or repelled from regions of high field strength according to the material dependent "Clausius-Mosotti" factor[22]. Since the dielectrophoretic force is proportional to the field gradient squared[22], not only does the accumulation of the nanoparticles indicate the presence of field gradients, but the number trapped also reflects its strength (see Supplementary Information, Eq.S2,S6-7).

The relationship between the position of the nanosensors and the applied voltage is obtained by recognizing that trapping will occur when the attracting dielectrophoretic potential exceeds the thermal potential (Brownian motion) such that.

$$V_{RMS}^2 \sim \frac{C}{\nabla \chi \left( x_1, x_2, x_3 \right)}; C = \frac{\Delta k_B T L^2}{\pi \varepsilon_f b^3 \operatorname{Re}\left\{ K_{CM} \right\}} \tag{1}$$

where $\nabla \chi$ is the derivative of the induced potential around the colloid, $k_B$ is the Boltzmann constant, $T$ is temperature, $L$ is the distance between electrodes, $\varepsilon_f$ is the permittivity of the fluid, $b$ is the nanosensor radius, $\operatorname{Re}\left\{ K_{CM} \right\}$ is the real part of the Clausius-Mosotti factor and $\Delta$ represents the ratio of the dielectrophoretic to thermal potentials[23] left here as a single fitting parameter. For the current experimental setup, $L = 120\,\mu m$, $k_B = 1.38 \times 10^{-23}\, m^2 kg s^{-2} K^{-1}$, $T = 298K$, $\varepsilon_f = 80 \cdot 8.85 \times 10^{-12}\, F/m$, $\mathrm{Re}\{K_{CM}\} \sim 1$ (high frequency) such that $C = 8\Delta$.

Experimentally, it has been found that Janus particles tend to be attracted to the wall and translate parallel thereto[24,25]. Qualitative examination of the 3D trapping around a stagnant Janus particle adjacent to a wall (Figure 1 b-d) illustrates that in the current setup, the strongest gradients are formed in the nanometer sized gap between the particle and wall (Figure 1d), with smaller gradients also observed at the upper interface between metallic and dielectric hemispheres (Figure 1c).

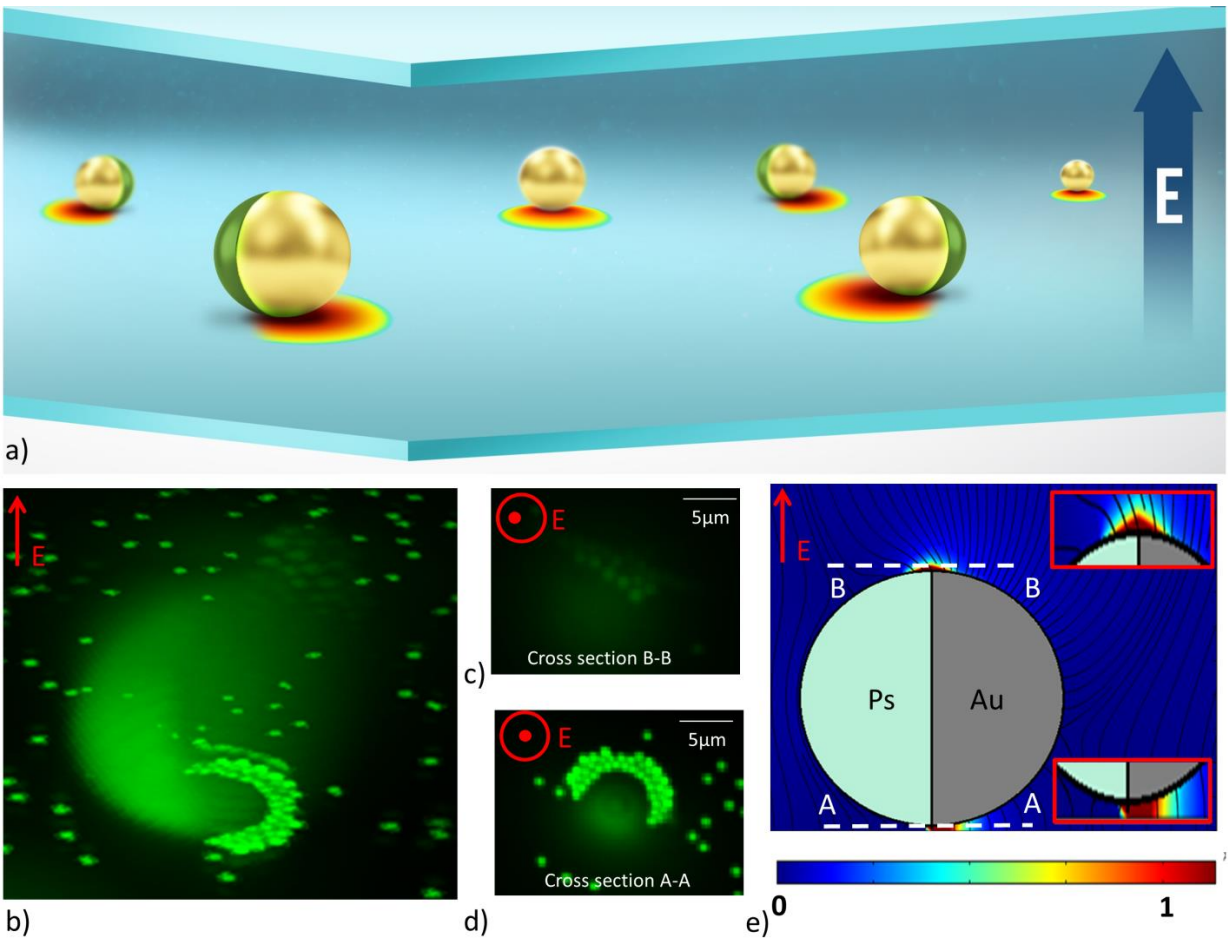

**Figure 1: Colloidal gradients** a) Schematic of the general concept of mobile microelectrode; localized gradients are induced around polarizable surface such that the spatial distribution of the field gradient in the system may be varied by manipulating the position of the Janus sphere, b-d) Experimental observation of the formation of field gradients around a Janus particle: b) 3D isometric image of trapped nanosensors - indicating regions of high electric field strength - generated from superposition of multiple images taken along the height of the particle; c) view of trapping at focal plane directly above the Janus sphere; d) view of trapping directly below the particle adjacent to the substrate; e) 2D COMSOL simulation of expected gradients around the Janus sphere.

The precise distribution of the electric field, $\nabla \chi$ induced around the mobile floating electrode is dependent on the geometry and material properties of the particle. Envisioning the polarized colloid as a source, one can theoretically derive the induced distribution for any geometry in terms of a multipole expansion, which decays to zero in the far-field (satisfying the Laplace equation) and where the magnitudes of the included terms are prescribed by the electrostatic boundary conditions at the colloid surface. When the colloid is in close proximity to the wall, it is necessary to also consider the images of the multipole expansion, since the interaction of the induced field with the channel boundaries can dominate - as in the current setup where gradients below the Janus sphere are significantly larger than those on top (Figure 1e).

For a metallodielectric Janus sphere, a good approximation of $\chi$ can be obtained by considering the sum of a point dipole and multipole, where the latter accounts for the symmetry-breaking (see Supplementary Information). Noting that maximal gradients will occur in the $x_2$ direction ($\nabla \chi \sim \partial \chi / \partial x_2$), and considering the limiting case of a particle in close proximity to the wall, $\left( h/a \sim 1 \right)$ (Figure 2a), the variation of the induced field magnitude along the line A-A is plotted to leading order in Figure 2b with the corresponding density plot over the cross-section A-A in Figure 2c. In order to parallel the experiments, the position of A-A $\left( x_2 = b \right)$ is chosen to coincide with the nanosensor center. We observe that maximum gradients are located close to the metallic hemisphere, and the field rapidly decays with distance from the colloid. It is noted that the finite values at the dielectric hemisphere are an artefact of the consideration of only the leading order dipole and multipole. Incorporation of higher orders would asymptotically reduce the induced field for $x_1 > 0$ to zero.

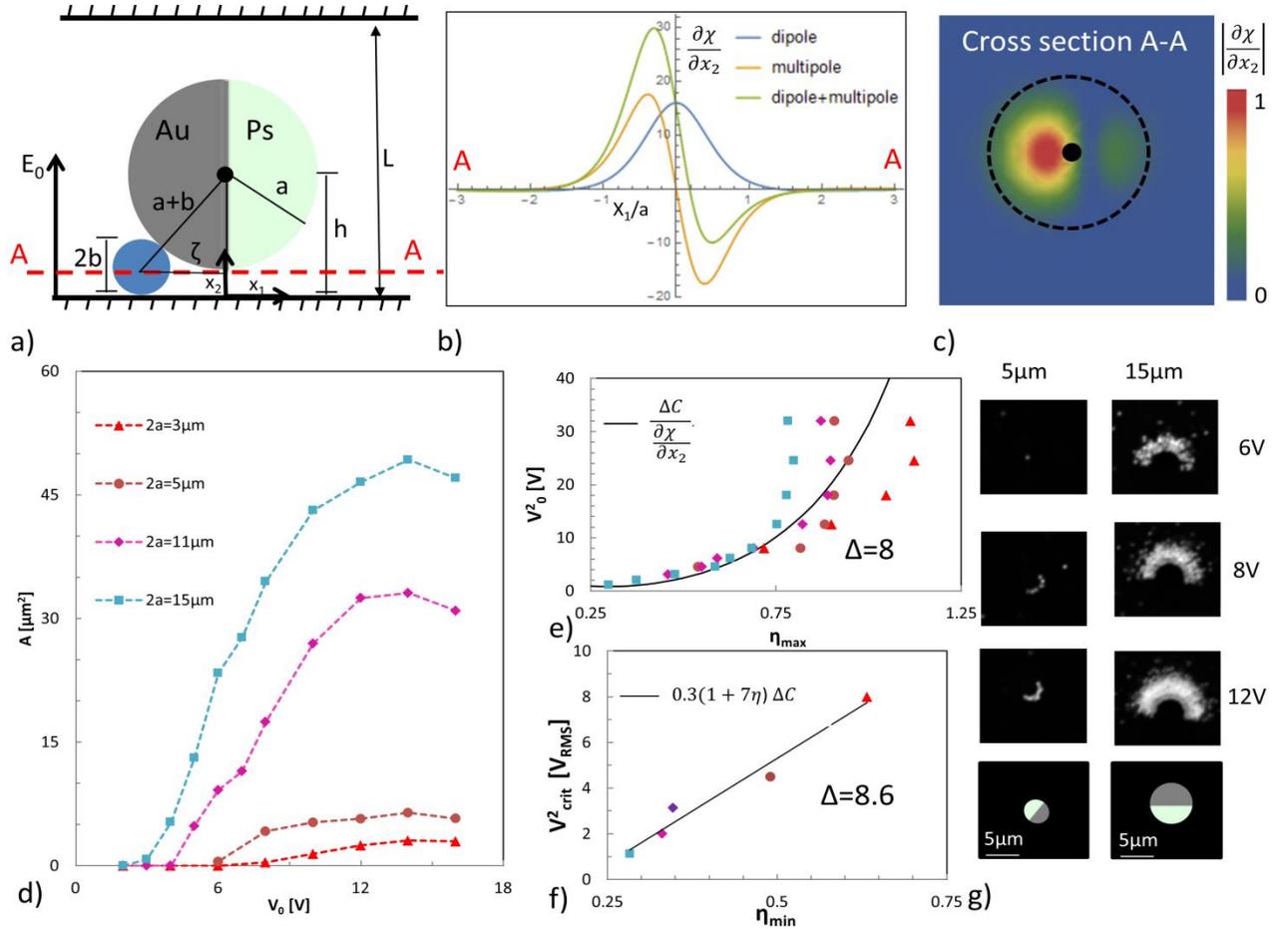

**Figure 2: Distribution of induced field around a Janus sphere** a) Schematic illustrating geometry of a Janus sphere, radius $a$ adjacent to a wall with a smaller nanosensor (radius $b$) trapped adjacent to its surface; b) Non-dimensional plot (normalized with respect to $a$) of the expected distribution of the electric field in the $x_1$ direction, along the center plane of the nanosensor $(b/a = 0.02)$ when only the dipole term is considered (uniform sphere), the multipole (symmetry breaking contribution) and the superposition of dipole and multipole corresponding to the distribution around a Janus sphere; c) Density plot of the absolute value of normalized electric field distribution beneath the Janus sphere along the cross section A-A. Dotted circle indicates radius of the Janus sphere and solid dot indicates center of the JP; d) Variation of area of trapped fluorescent nanosensors with applied voltage for 300nm sensors around Janus spheres 3,5 ,11 and 15μm in diameter; e) Scaling of the increased trapping area as a function of the applied electric field $\eta_{max}^2 = \dfrac{x_{1,max}^2}{a^2} \sim \dfrac{2A + \pi\zeta^2}{\pi a^2}$; f) Scaling of the minimum critical voltage required for onset of trapping with the electric field for varying Janus particles sizes $\eta_{min} = |\zeta/a| = 2\sqrt{b/a}$; g) Microscope images of trapping around Janus spheres 5 (left column) and 15μm (right column) in diameter at for applied fields of 6,8 and 12V with a schematic (bottom row) indicating orientation of the JP.

The theoretical model is verified experimentally by measuring the dependence of the area covered by the trapped nanosensors on the applied voltage (Figure 2d). As the voltage is increased, the critical field corresponding to the dominance of DEP over thermal forces occurs farther from the center of the JP. Since the area, $A$, of trapped nanosensors (fluorescent region in Figure 2g) plotted in part d of Figure 2 is approximately hemispherical, we can express $A$ in terms of the nondimensional parameter $\eta = \left| \dfrac{x_1}{a} \right|$ such that $\eta_{max}^2 = \dfrac{x_{1,max}^2}{a^2} \sim \dfrac{2A + \pi \zeta^2}{\pi a^2}$. Here $\zeta$ corresponds to the geometrically dictated minimum distance from the center of the particle that can be reached by the tracer (Figure 2a). In the limit of $h/a \sim 1$, $\zeta = 2\sqrt{ba}$, so that $\eta \big|_\zeta = \eta_{min} = 2\sqrt{b/a}$.

In Figure 2e, we demonstrate that the relationship between $\eta$ and $V$ is in good agreement with the theoretical curve obtained for $\partial \chi / \partial x_2 = -2 \left( \dfrac{P_1 \left( 2 - \eta^2 \right)}{\left( 1 + \eta^2 \right)^{5/2}} + \dfrac{6 P_2 \eta \left( 4 - \eta^2 \right)}{\left( 1 + \eta^2 \right)^{7/2}} \right)$ (Supplementary Information, Equation S8); where the saturation of $\eta_{max}$ at slightly lower voltages than theoretically predicted is likely due to $h/a > 1$ and may also be related to particle-particle interactions. For low voltages, a simpler scaling can be obtained using the linear approximation $C / \left( \partial \chi / \partial x_2 \right) \sim 0.3 \left( 1 + 7\eta \right)$ (Supplementary Information, Eq.S10); yielding good agreement for the critical applied voltage at which the first nanosensor is trapped at $\eta_{min}$. In both cases, similar values of $\Delta = 8, 8.6$ are obtained; corresponding well with the general rule of thumb that requires the DEP potential to exceed the thermal by an order or magnitude for onset of trapping[23].

In the sequel, we highlight one practical application of the portable microelectrodes, which demonstrates a number of the advantages offered by colloidal based gradients vis-a-vis the current state of the art, top-down processes. Noting that under the application of a uniform field, not only do Janus particles produce field gradients strong enough to assemble secondary elements at its surface, but that they also move as active colloids[15], we demonstrate an all-in-one cargo-carrier, capable of *selectively* accumulating and transporting a target to a secondary location where it can be released on demand for collection or further analysis (Figure 3a-b, Supplementary Information, Movie S1). The only control parameter required is the frequency of the applied field; which determines both the direction of the JP[15,25] and distribution of the assembly as well as and whether the targets are attracted (accumulation) or repelled (release) from its surface.

At low frequencies, Janus particles translate under induced-charge electrophoresis (ICEP) where strong hydrodynamic flow around the metallic hemisphere propels the Janus particle with its dielectric hemisphere foreword. The combination of the induced-charge electroosmotic flow, which injects from the bottom of the Janus sphere and ejects at the midplane[14] and the location of the tracers at the aft of the particle results in the concentration of the cargo around the midline of the JP (see Figure 3d,I, ). At high frequencies, there is no hydyrodynamic flow and the Janus particle moves with its metallic hemisphere forward, likely due to the unbalanced gradients of the applied field[25].In the high frequency case, where there is no hydrodynamic flow and cargo is located fore of the particle, the distribution is hemispherical and remains a monolayer at moderate voltages(Figure 3d, II-IV, Supplementary Information, Movie S2).

The selectivity arises from the fact that all materials exhibit a unique frequency response dependent on their geometric and material properties (complex permittivity)[21], such that by aligning the frequency of the applied field to coincide with the pDEP response of a target and nDEP response of a secondary contaminant within the solution, we may dynamically select a desired target from a mixture (Figure 3c-d). This selective mode is highlighted in green (III in Figure 3c-d), where the forcing frequency of 750kHz is chosen to coincide with pDEP of the 300nm target which accumulates at the surface of the JP while the 1µm "contaminant" is repelled by nDEP .

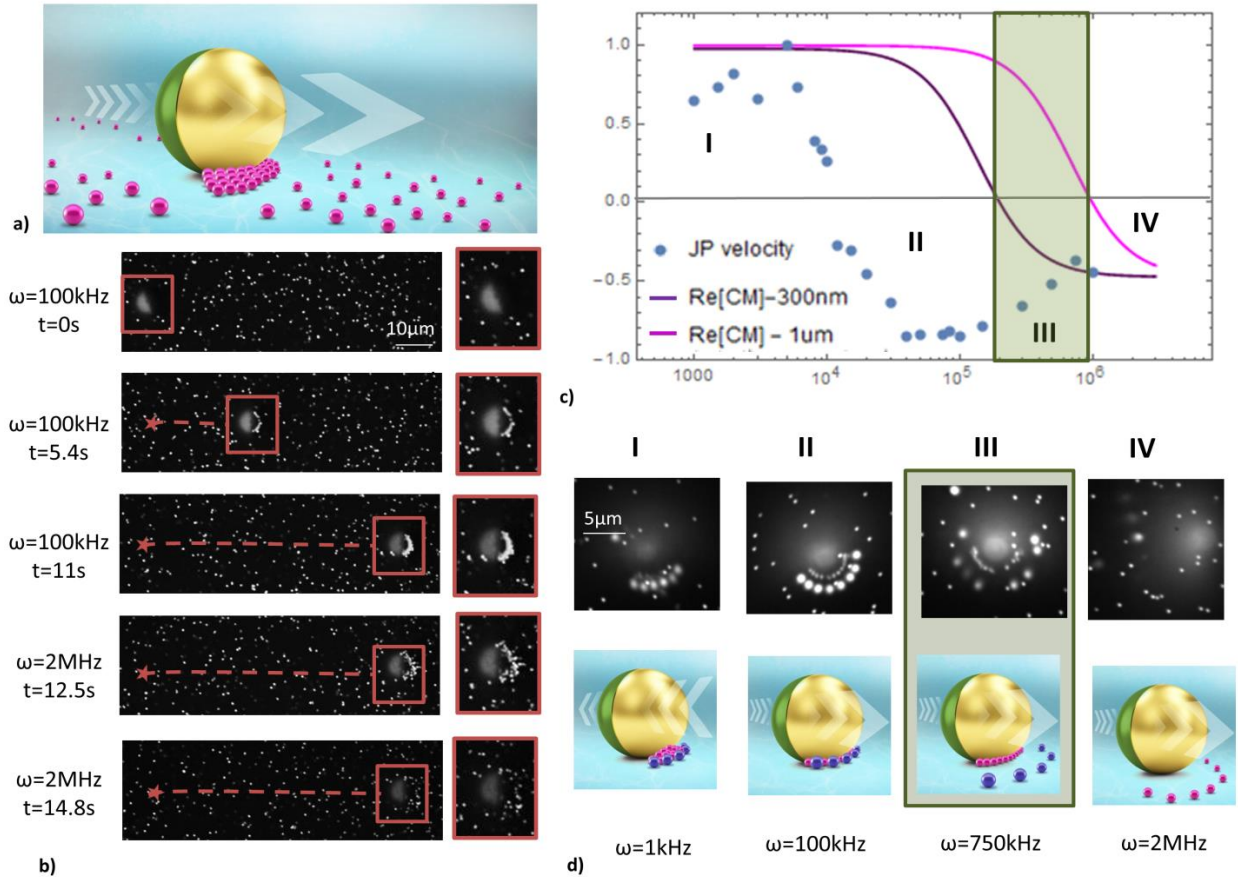

**Figure 3: Selective cargo assembly and transport** a) Schematic illustrating concept of cargo transport; Janus particle picks up cargo as it translates and transports it to a secondary location b) Microscope images showing accumulation, transport and release of 300nm target carrier by 15µm JP controlled by the frequency of the applied electric field, c-d)Precise control of the frequency of the applied field enables selective transport (green highlighted region). The desired operating frequency is chosen to align with positive DEP of target (magenta) and negative DEP of the contaminant (purple). The frequency of the applied field also determines the mode of transport –whether by ICEP with Ps hemisphere forwards or sDEP with metallic hemisphere forwards. c) Operating conditions can be determined by plotting the frequency dependent mobility of JP carrier (blue data points) and the real part of the Clausius-Mosotti factor of target and contaminant (magenta and purple curves). Four frequency domains are observed – denoted by Roman numerals I-IV each of which correspond with microscope images (d). I) At low frequencies, target and contaminant undergo pDEP (non-selective trapping) while JP translated forward under ICEP; II) Janus particle reverses direction but both target and contaminant still undergo pDEP; III) Selective trapping – target undergoes pDEP while contaminant undergoes nDEP; IV) Frequency aligned with nDEP of target for release

In contrast to contemporary carriers utilizing magnetic[20,21] or chemical functionalization[26] as a means of attraction, the current configuration, which relies on dielectrophoresis to trap the target, can provide selective transport for a wide range of media (organic and inorganic), with the flexibility to dynamically select and release the target by altering the frequency of the applied field. At the same time, when compared to standard microfluidic dielectrophoretic devices, which require complex (costly) patterning to yield gradients sufficiently strong to compensate for the decrease of the dielectrophoretic force with the radius cubed, the colloidal cargo carrier is best suited to nanotargets since $\eta_{min} = 2\sqrt{b/a}$. The motion of the particle also simplifies the system since it eliminates the necessity of a secondary convection mechanism (e.g., pressure driven flow) to bring the target to the region of high field strength[10,27]

Thus, within this single example, we highlight a number of the advantages to be expected when we shift from top-down to bottom up processes and smart devices including increased energy efficiency (trapping nanotargets at moderate voltages), simplified fabrication processes and the flexibility to reconfigure the system to multiple tasks and environments. It is emphasized however that the general concept of field gradients built from the bottom up, need not be limited to those traditionally associated with either active colloids or dielectrophoresis, but may be combined with directed motion towards applications such as bottom-up building of smart materials where the frequency dependent selectivity could be combined with immobile (homogeneous) colloids to develop complex building blocks for self-assembly.

**Methods**

### 1. Janus particle and solution preparation

Janus particles 3, 5, 10 and 15μm in diameter were manufactured by coating green polystyrene spheres (Fluoro-max) with 10nm Cr followed by 30nm Au according to the protocol outlined in Ref.25. To enable selective illumination, the 300nm Ps particles (Fluoro-max) used in Figures 2, and 3 are red while the 720nm and 1μm in Figures 1 and 3 respectively are green. Particles were rinsed three times with DI water, to which a small amount of nonionic surfactant (Tween 20 (Sigma Aldrich)) was added in order to minimize adhesion to the ITO substrate before being injected into the microfluidic chamber via a small hole at the upper substrate, drilled expressly for this purpose. Based on the size of the polystyrene particles, their concentration ranged between 0.1-0.01% (w/v).

2. **Experimental set-up**

The experimental chamber consisted of a 120μm high silicone reservoir (Grace-Bio) sandwiched between an ITO coated glass slide (Delta Technologies) and an ITO coated coverslip (SPI systems) as illustrated in Ref.25. Two holes were drilled at the top of the chamber to ensure the chamber remained wetted by enabling the addition of electrolyte and colloids into the channel via manual pumping. These holes were surrounded by a reservoir, 2mm in height filled with solution. The AC electric forcing was applied using a signal generator (Agilent 33250A) and monitored by an oscilloscope (Tektronix-TPS-2024).

3. **Microscopy and image analysis**

Observation of the distribution of Polystyrene colloids around the Janus sphere (Figures 1-3 were performed on a Nikon Eclipse Ti-E inverted microscope equipped with Yokagawa CSU-X1 spinning disk confocal scanner and Andor iXon-897 EMCCD camera. The chamber was placed with the cover-slip side down and images were magnified with x60 oil immersion lens. Green and red particles were illuminated with lasers of wavelength 488nm and 561nm respectively. To obtain the frequency dispersion in Figure 3, the microfluidic chamber was observed under a Nikon TI inverted epi-fluorescent microscope, fitted with a x10 lens and recorded on a Andor

Neo sCOMS camera for a minimum of 30s at a rate of 5 frames/s. Particle velocities were extracted by tracking particle displacement using PredictiveTracker function in Matlab (https://web.stanford.edu/~nto/software.shtml)) and averaging the velocity over the number of mobile particles. Accuracy of the code was initially verified by comparing the velocities obtained using this method with manual measurement in Image J.

### 4. Numerical simulations

The numerical simulation used to qualitatively verify the presence of asymmetric electric field gradients arising from the proximity of a Janus sphere near a conducting wall was performed in COMSOL™ 4.2. A simple 2D geometry was used to model the experimental setup; consisting of a rectangular channel, $80\mu m$ height and $200\mu m$ wide, with a $15\mu m$ diameter circle placed $300nm$ above the substrate (this height is based on the observation that the 300nm tracers do not pass below the JP). The electrostatic equations are solved in the rectangular domain, with the following boundary conditions. At the lower substrate ($y$=0) a voltage of 5V is applied while the upper wall is grounded. The edges of the channel are given an insulating boundary condition. The Janus sphere is modelled by applying floating electrode and insulating boundary conditions at the metallic (right) and dielectric (left) hemispheres respectively.

**Acknowledgements**

The authors would like to acknowledge US-Israel Binational Science Foundation Grant 2009371, Israel Science Foundation 1945/14 and the RBNI and Gutwirth graduate fellowships. The JP preparation was possible through the financial and technical support of the Technion RBNI (Russell Berrie Nanotechnology Institute) and MNFU (Micro Nano Fabrication Unit). We also thank Dr. Sinwook Park for technical assistance.

**Author Contributions**: AB performed/supervised experiments, image and data analysis, numerics, developed scaling analysis and compiled the manuscript. TB performed experiments and image analysis. TM was responsible for theoretical derivation.GY supervised planning and execution of experiments and assisted in scaling analysis and numerics. All authors contributed to preparation of the manuscript.

**Competing Financial Interest:** The authors declare no competing financial interest.